\documentclass[pre,aps,twocolumn,superscriptaddress,floatfix]{revtex4-2}
\usepackage[english]{babel}
\usepackage{float}
\usepackage{color}
\usepackage{graphicx}
\usepackage{dcolumn}
\usepackage{bm}
\usepackage[normalem]{ulem}
\usepackage{xcolor}
\usepackage{amsmath}

\newcommand{\keff}{K_{\mathrm{eff}}}
\newcommand{\heff}{H_{\mathrm{eff}}}
\newcommand{\pir}{p_{\mathrm{Ir}}}

\newcommand{\hk}{H_{\mathrm{K}}}
\newcommand{\hd}{H_{\mathrm{d}}}
\newcommand{\td}{T_{\mathrm{d}}}
\newcommand{\ms}{M_{\mathrm{S}}}
\newcommand{\sigmaC}{\sigma_{\mathrm{C}}}

\newcommand{\hdmi}{H_{\mathrm{DMI}}}
\newcommand{\hx}{H_{\mathrm{x}}}
\newcommand{\vd}{v_{\mathrm{d}}}
\newcommand{\hcoer}{H_{\mathrm{Coer}}}
\newcommand{\hsat}{H_{\mathrm{Sat}}}
\newcommand{\area}{\mathcal{A}}
\newcommand{\perimeter}{\mathcal{P}}
\newcommand{\vldmi}{v^{L}_{\mathrm{DMI}}}
\newcommand{\vrdmi}{v^{R}_{\mathrm{DMI}}}
\newcommand{\tsamp}{t_{\mathrm{S}}}
\newcommand{\mmin}{n}

\begin{document}

\title{Effect of Ir growth pressure on the domain wall dynamics in Ta/Pt/Co/Ir/Ta stacks}

\author{P. Domenichini}
\affiliation{INENCO, CONICET. Av. Bolivia 5150 (A4400), Salta Capital, Salta, Argentina}
\affiliation{Departamento de Física, Fac. de Cs. Exactas, Univ. Nac. Salta. Av. Bolivia 5150 (A4400), Salta Capital, Salta, Argentina}

\author{Jeffrey A. Brock}
\affiliation{Center for Memory and Recording Research, University of California - San Diego,
9500 Gilman Drive, La Jolla, CA, 92093, USA}

\author{J. Curiale}
\affiliation{Instituto Balseiro, Univ. Nac. Cuyo - CNEA, Av. Bustillo 9500 (R8402AGP), S. C. de Bariloche, Río Negro, Argentina.}
\affiliation{
Instituto de Nanociencia y Nanotecnología (CNEA-CONICET), Nodo Bariloche, Av. Bustillo 9500 (R8402AGP), S. C. de Bariloche, Río Negro, Argentina}
\affiliation{Gerencia de Física, Centro Atómico Bariloche (CNEA), Av. Bustillo 9500 (R8402AGP), S. C. de Bariloche, Río Negro, Argentina.
}

\author{A. B. Kolton}
\affiliation{Instituto Balseiro, Univ. Nac. Cuyo - CNEA, Av. Bustillo 9500 (R8402AGP), S. C. de Bariloche, Río Negro, Argentina.}
\affiliation{Gerencia de Física, Centro Atómico Bariloche (CNEA), Av. Bustillo 9500 (R8402AGP), S. C. de Bariloche, Río Negro, Argentina.
}

\date{\today}

\begin{abstract}

The dynamical response of magnetic domain walls in ultrathin magnetic films to external magnetic fields is determined not only by the composition and thickness of the layers but also by the growth conditions. Growth conditions can induce significant structural changes to the layers and at the interfaces between them, affecting in particular the dynamics of domain walls, their mobility, elastic tension, and the pinning forces acting on them. In this work, we focus specifically on the effect of Ir layer growth pressure in Ta/Pt/Co/Ir/Ta ultrathin multilayered films. Measurements of the DC magnetic properties, domain wall velocity, and domain morphology in the creep regime for both constant and alternating field pulses were performed for a batch of samples where the Ir layer was grown at different pressures. We find that the saturation magnetization, the effective anisotropy constant, and the domain wall surface tension grow with increasing pressure and saturate at a threshold pressure, while the Dzyaloshinskii-Moriya field and the strength of the disorder remain practically unaltered over the range of pressures considered.

\end{abstract}

\pacs{Valid PACS appear here}

\maketitle

\section{ Introduction}\label{sec:Intro}

The special properties that emerge when one or more characteristic dimensions of a magnetic system are reduced to the nanometer scale have attracted significant attention in recent years due to their promising technological applications~\cite{KUMAR20221}. Among the most studied examples of such systems are ultra-thin ferromagnetic films, which consist of multilayer stacks with layers made of different materials of nanometric thicknesses. These systems are particularly sensitive to the orientation and strength of magnetization \cite{Art-NatNanoTec8-Fert, Art-JMMM568-Corodeanu, Art-JMMM383-Awano, Art-Science309-Allwood, Art-JPD53-Road}, and thus potentially attractive for novel devices. However, their development entails complex nanofabrication methods.
Both the film composition and growth conditions have been shown to strongly influence the magnetic properties, type of domain wall (DW) stabilized, and DW propagation ~\cite{Art-JPD54-Quinteros,Art-PRB104-Burgos, Art-JAP106-Posth}. Strikingly, despite their sensitivity, magnetic thin films with perpendicular anisotropy have also become a model experimental system for studying the universal dynamical properties of one-dimensional extended elastic interfaces in random media ~\cite{Wiese2022}. This universality provides a convenient phenomenological framework for exploring the effect of growth conditions on DW dynamics by focusing the attention on the constitutive equations of a few experimentally accessible transport coefficients.

The dynamics of magnetic DWs are mainly determined by the interplay between elasticity, external fields, thermal fluctuations, and quenched disorder. The surface tension and the type and strength of the pinning forces acting on a domain wall critically depend on microscopic properties, including bulk and surface anisotropies, asymmetries, spatial heterogeneities and defects, dipolar interactions, exchange interactions, and the Dzyaloshinskii-Moriya interaction (DMI). The presence of DMI in ultra-thin films broadens the possibilities of these systems further, owing to its relevance in generating intriguing magnetic textures such as chiral domain walls and skyrmions~\cite{Kuepferling2023}. These small magnetic textures hold significant potential for application in the next generation of digital processing and recording techniques based on spintronics, which aim to reduce power consumption while enhancing capacity, processing speed, and robustness~\cite{Art-NatNanoTec8-Fert}.

Among the film stacks that have been studied are Pt/Co films with top layers composed of heavy metals (HMs) such as Pt, Ni, Ta, Ir, among others \cite{Kuepferling2023, Art-PRB91-Lavrijsen, Art-PRB95-Wells, Art-PRB90-Hrabec,Art-JofPDAP-Belmeguenai, Art-AppSurSci543-Kolesnikov, Art-PRB99-Shahbazi}. The effects of changing Ir thickness in Ta/Pt/Co/Ir/Ta stacks on the resulting DMI and domain wall dynamics were recently analyzed in Ref.~\cite{Art-PRB99-Shahbazi}, revealing substantial changes in DMI strength. Another interesting growth modification is to use different Ar pressures when growing the top layer of the system. There are few studies that demonstrate a significant change in the dynamic properties of the samples \cite{Art-PRB91-Lavrijsen, Art-AppSurSci543-Kolesnikov}, but particularly none in Pt/Co/Ir films. In all of these works the magnetic characteristics of the samples are analysed in the presence of a constant magnetic field (DC field), but the response to an alternating magnetic field (AC) was not explored. Recently, it was shown that the AC response yields valuable information about domain walls and pinning properties. These properties are complementary to those obtained by using DC fields, as demonstrated in Refs. \cite{Art-PRB99-Domenichini,Art-PRB103-Domenichini}.

In this paper, samples of SiO$_2$//Ta(5)/Pt(3)/Co(0.8) /Ir(1)/Ta(2) (thicknesses in nm), are investigated. Samples were grown via DC magnetron sputtering with different Ar pressures used during the growth of the Ir layer; we refer to the Ir growth pressure as $\pir$. All samples have the same nominal thickness for each layer. We analyze the changes observed in the hysteresis loops, static micromagnetic properties, domain wall velocities under a DC/AC field, and DMI intensity through asymmetric domain growth in presence of in-plane and out-of-plane fields.

\section{Properties of Interest}
\label{sec:properties}

In perpendicularly magnetized ultrathin films with bubble-like domains, the study of the domain wall (DW) velocity as a function of a constant and uniform magnetic out of plane field $H$ is a standard probe to study dynamical properties of DWs.
The velocity-field characteristics $V(H)$ thus obtained reveals several dynamical regimes with interesting universal features \cite{Art-PRB98-Jeudy, Art-PRL80-Lemerle, Albornoz2021}.
These universal features manifest at low velocities, in the creep~\cite{Art-PRL80-Lemerle} and depinning \cite{Albornoz2021} regimes, for $H\ll \hd$ and $H \sim \hd$ respectively, with $\hd$ the depinning field.
The most robust universal property is unveiled in the low-field regime with $T>0$ and $H \ll \hd$ which is well  characterized by the creep law 
\begin{equation}
    V(H) = V_0 \exp ( -\alpha (\mu_0 H) ^{- \mu} + \beta),
    \label{eq:VelCreep}
\end{equation}
where $V_0$ is a characteristic velocity, $\alpha$ and $\beta$ are material and temperature dependent but field-independent parameters and $\mu_0$ the vacuum permeability. This law describes a thermally activated motion of thermal nuclei 
over typical energy barriers $U_{\tt opt}\sim \alpha T (\mu_0 H)^{-\mu}$. 
The creep exponent is predicted to be $\mu=1/4$ for one-dimensional interfaces in random-media with the so-called ``random-bond'' type of disorder \cite{nattermann_creep_full,chauve2000,Ferrero2013,Ferrero2021}. The exponent $\mu=1/4$ in particular is consistent with a large number of measurements performed in ultrathin films of different materials~\cite{Art-APL112-Quinteros, Art-PRB104-Burgos, Art-APL97-Kanda,gorchon2014,jeudy2016,Art-PRL99-Metaxas,grassi2018,Art-PRB99-Shahbazi,Kuepferling2023}. These experiments confirm the predictions of elastic interface models, the effective dimensionality of the DWs and the type of disorder present in the samples. Indeed, a ``random-field'' type of disorder, a non-local elasticity or a different DW dimensionality is also well predicted by Eq.(\ref{eq:VelCreep}) but with a clearly distinguishable $\mu \neq 1/4$~\cite{Ferrero2021}. The creep formula of Eq.(\ref{eq:VelCreep}) thus offers a robust phenomenological framework to analyze the effect of growth conditions in the pinning and elastic properties of DWs. Furthermore, $\alpha$ and $\beta$ can be predicted by the Larkin collective pinning theory \cite{blatter_vortex_review,nattermann2000,chauve2000}. Using Larkin collective pinning theory \cite{blatter_vortex_review,chauve2000} it is possible to connect $\alpha$ and $\beta$ with different parameters such as the DW surface tension $\sigma$, the saturation magnetization $\ms$, the DW width $\delta$ as well as with relevant pinning parameters such as the strength of the disorder $\Delta_0$ and the pinning correlation length $\xi$ (we describe and exploit these relations in Section \ref{sec:discussion}). All these parameters are in principle sensitive to growth conditions and in particular $\pir$. An example of the effect of growth pressure on Pt/Co/Pt samples is reported by Lavrijsen \textit{et al.} \cite{Art-PRB91-Lavrijsen}, where growth conditions are shown to strongly impact the DW velocity profiles of expanding bubbles under the simultaneous application of out-of-plane and in-plane magnetic fields.

In perpendicularly magnetized systems, by applying an in-plane magnetic field ($H _{\mathrm{x}}$) in addition to an the out-of-plane field $H$, asymmetric growth of bubble domains can be observed if DMI is present~\cite{Art-PRB95-Wells, Art-NM15-Jue, Art-PRB94-Lau, Art-PRB88-Je}.
In these systems, DWs grow with different speeds depending on their orientation relative to $H _{\mathrm{x}}$. When velocity is measured parallel to $H _{\mathrm{x}}$ a minimum is observed when $H _{\mathrm{x}} = \hdmi$,  
\begin{equation}
\min_{\hx} V(H,\hx)=V(H,\hdmi)    ,
\label{eq:hdmidef}
\end{equation}
with $V(H,\hx)$ the velocity-field characteristics under the tilted field, such that $V(H,\hx=0)\equiv V(H)$ of Eq.(\ref{eq:VelCreep}) in the creep regime. 
{
From $\hdmi$ we can estimate the  
DMI strength as
\begin{equation}
D=\mu_0 \hdmi \ms \delta   
\label{eq:DMIstrength}
\end{equation}
with $\delta=\sqrt{A/\keff}$ the DW-width, $A$ the exchange interaction, $\keff$ the effective anisotropy and $M_s$ the saturation magnetization~\cite{Gehanne2020}.}
An isotropic growth implies $\hdmi=0$, and thus, a negligible DMI. 
In our case, the measurement of $\hdmi$ for samples with different $\pir$ allows us to study not only the role of the interface engineering on the strength of the DMI but also to test theoretical predictions relating it to changes in the total surface tension $\sigma$, which we can experimentally estimate  as explained in the following.

Complementary information on the pinning and elasticity of DWs
can be obtained by applying a square wave of an out of plane magnetic field $H(t)$ to a compact domain \cite{Art-PRB99-Domenichini}. 
In the absence of any magnetic field, a micrometer sized domain in an ultrathin disordered film is typically trapped into a metastable state from which it can not escape within experimental time-scales, in spite of having a negative average curvature promoting its collapse. 
The applied pulses are hence chosen to be large enough to induce an AC-assisted curvature driven collapse of the DW in experimental time scales. This method exploits the approximation that the local response of a DW to an external field is given approximately by its DC velocity response $V(H_{\tt eff})$ in an effective local field $H_{\tt eff}=H(t)+C \kappa$, where $\kappa$ is the local signed curvature of the DW and $C = \sigmaC/2M_S$, with $M_S$ the saturation magnetization and $\sigmaC$ a phenomenological elastic  tension, expected to be of the same order of magnitude than the micromagnetic prediction $\sigma_0 \approx 4\sqrt{A \keff}$, with $A$ the exchange interaction and $\keff$ the effective anisotropy coefficients. Under this approximation and also assuming that the area oscillations are small compared to the total area it can be shown that at time $t=N\tau$ we have the effective area
~\cite{Art-PRB103-Domenichini}
\begin{equation}
     \Lambda(N) \equiv - \frac{\area(N)+\perimeter(N) \, V(H) \, (\tau /4)}{2 \pi V'(H) \tau} \approx \Lambda(0) + C N,
     \label{eq:Lambda}
\end{equation}
where $\area(N)$ is the average area of the domain, $\perimeter(N)$ is the average perimeter of the DW at the $N$-th pulse, 
$\tau$ the period of the square wave, 
$V(H)$ is the DC velocity-field characteristics and $V'(H)$ its derivative with respect to the AC field amplitude $H$.
Eq. (\ref{eq:Lambda}) generalizes the  exact and universal spontaneous decay $A_t = A_0 - 2\pi C m t$ expected for the area of any compact domain of initial area $A_0$ when the velocity response is simply $V(H)=mH$, with $m$ the DW mobility, a situation typically realized in the absence of pinning at relatively small or large fields.
Numerical simulations of an elastic DW in presence of an AC field show that in the absence of disorder Eq. (\ref{eq:Lambda}) is very accurate.  In the presence of disorder however, it holds only for small $N$, provided the initial condition is not strongly correlated with the underlying disorder. 
The later condition is typically realized just after nucleation of a compact domain by a short and high magnetic field pulse applied in a saturated magnetization background. 
We can thus use Eq.(\ref{eq:Lambda}) to extract $C$ by only monitoring $\area(N)$ and $\perimeter(N)$, both accessible from images.
Furthermore, as long as we have access to $M_S$, we can compute $\sigmaC$.

For larger $N$, the prediction $\Lambda(N)-\Lambda(0) \sim  C N$ is expected to fail due to the roughening of the DW. 
For nearly circular initial domains, DW roughening can be monitored in the same experiment by using the quadratic mean displacements of the DW with respect to a perfect circle with a radius equal to the angularly averaged time-dependent radius of the deformed DW loop. 
If $r(\theta,N)$ is an univalued function that describes the DW in polar coordinates from the nucleation center in the $N$-th pulse, we can define the $N$ dependent roughness as
\begin{eqnarray}
    \langle u^2 \rangle = \langle r^2\rangle_{\theta} - \langle r \rangle_{\theta}^2 
\end{eqnarray}
where $\langle \mathcal{F} \rangle_\theta \equiv (2\pi)^{-1}\int_0^{2\pi} \mathcal{F}d\theta$.
The quantity $\langle u^2 \rangle$ is expected to grow with $N$ and to scale as a generic roughening process $ \langle u^2 \rangle  \sim N^{ \gamma }$, with $\gamma$ a growth exponent~\cite{barabasi_stanley_1995}.

To complete the set of properties just described we also analyze the saturation magnetization $\ms$ and the anisotropy field $\hk$. All the above properties are measured as a function of $\pir$.

\section{Experimental Details} \label{Sec:Exp}
\label{sec:experiment}

We studied ultra-thin films with perpendicular magnetic anisotropy (PMA) grown by DC magnetron sputtering on thermally oxidized SiO$_2$ substrates.
The structure of the samples was SiO$_2$//Ta(5)/Pt(3)/Co(0.8)/Ir(1)/Ta(2) (the nominal thickness of each layer is specified in nm between parentheses). All layers were grown using Ar pressure of $3$ mTorr, except for the Ir layer, which was grown with different pressures $p_{\text Ir}$, ranging from $3$ to $25.5$ mTorr, while maintaining a constant nominal thickness for the Ir layer.
All layers were grown using a sputtering power of 50 W.
Before deposition, the sputtering rate of each material was calculated by growing a film for ten minutes and performing X-ray reflectivity (XRR) measurements to determine the thickness of the deposited material. A separate growth rate was calculated for each $\pir$ employed. By performing XRR and rocking curve measurements of the multilayer structures, the total thickness of the films and the roughness and elemental intermixing were assessed.

The magnetic properties of each sample were characterized using two techniques: A superconducting quantum interference device (SQUID) to measure the static magnetic properties, and a homemade polar magneto-optical Kerr effect (PMOKE) microscope to image the magnetic domain morphology of the samples and characterize the dynamics and morphology of domain walls.
Hysteresis loops as a function of the in-plane and out-of-plane magnetic field were measured using SQUID at room temperature. This allowed us to obtain the saturation magnetization values $\ms$, the effective anisotropy field $\hk$ to calculate the effective anisotropy constant $\keff = \mu_0 \hk \ms /2 $ \cite{Book-IntroMag-Cullity}, and confirm that all the samples present a dominant PMA. Furthermore, to analyze the morphology of the domains during reversal, images were taken using PMOKE microscopy. To build the PMOKE hysteresis loop we applied $50$ ms square pulses increasing/decreasing amplitudes in steps of $\mu_0 \Delta H$=$0.5$ mT and images were taken between pulses.

The DW dynamics were characterized by PMOKE using three different and complementary approaches. 
{
All protocols start by nucleating a compact magnetic domain
by applying out-of-plane magnetic field pulses of magnitudes on the order of $20$ mT, with a duration of $5$ ms, opposed in direction to a previously out-of-plane magnetically saturated background.
\begin{enumerate}
    \item[(i)] In the first, DC characterization, we apply a sequence of out-of-plane magnetic field square pulses of amplitude $H$ and different durations to grow the nucleated domain in successive steps (see for instance inset of Fig.\ref{fig:Vel-MOKE}(b)). As described in Ref.\cite{Art-PRB99-Domenichini}, the sequence of average DW displacements for a given pulse duration is used to estimate the average velocity $V(H)$ for $4\; \rm{mT } \leq \mu_0 H \leq 35\; \rm{mT }$, and in particular to determine the constants $\alpha$ and $\beta$ associated with the creep regime through Eq.(\ref{eq:VelCreep}).
    \item[(ii)] In the second, DC characterization, we apply a sequence of identical out-of-plane magnetic field square pulses of amplitude $H$ to grow the nucleated domain in steps, in the presence of a constant in-plane magnetic field $\hx$. We thus obtain $V(H,\hx)$, for $-300$ mT $\leq \mu_0 \hx \leq 300$ mT and $\mu_0 H = 100\; \rm{mT}$, and $\hdmi$ determined using Eq. \eqref{eq:hdmidef}.
    \item[(iii)] In the third AC characterization a pure ({\it i.e.} without a DC component) out-of-plane square-wave magnetic field is applied. Every period of the square-wave, of duration $\tau = 50$ ms, consists of a positive square-pulse of duration of $25$ ms immediately followed by a negative square-pulse of the same amplitude and duration. The amplitude of the square-wave is fixed so to always produce a fixed $6 \;\mu\rm{m}$ displacement in the first half-period positive pulse.
    Since the curvature of the DW plays a relevant role in this protocol, we ensure that the initially nucleated domain has a fixed $\sim 40 \;\mu{\rm m}$ radius, and then acquire images every period to determine the area $\area$ and perimeter $\perimeter$ of the domain, and then the $\Lambda$ function from Eq. \eqref{eq:Lambda} as a function of the period number $N$, to finally estimate $C$.
\end{enumerate}
}

\section{Results}
\subsubsection{Structural characterization}
\label{sec:RX}

\begin{figure}[]
  \centering
    \includegraphics[scale=0.58]{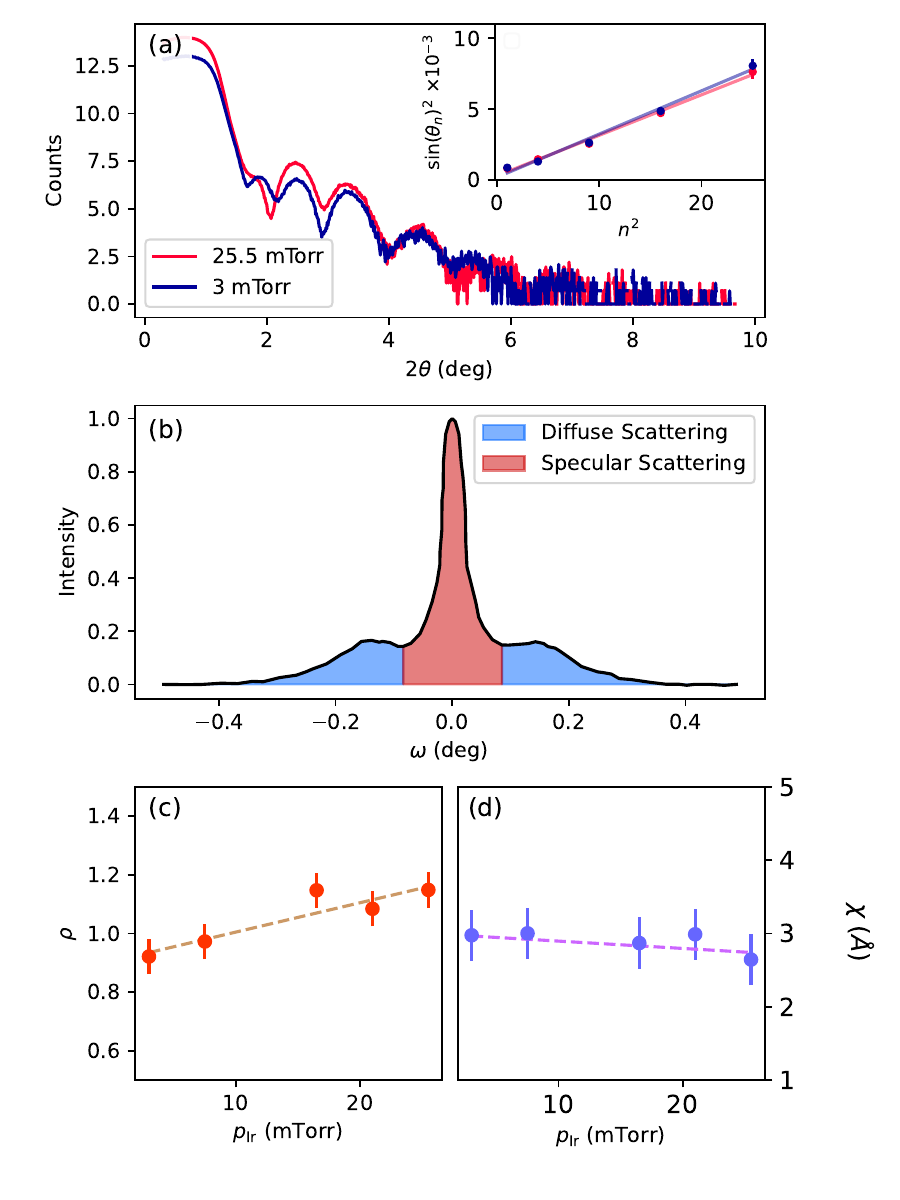}
    \caption{(a) Low angle X-ray reflectivity data for Pt/Co/Ir structures deposited with $\pir = 3$ and $25.5$ mTorr. The inset shows the linear fits using Eq. (\ref{Eq:RX}) and the obtained values for $\tsamp$. (b) Rocking curve for $\pir = 3$ mTorr, acquired around the first Kiessig fringe. Different colors corresponds to diffuse and specular scattering. Global roughness parameter $\rho$ and $\chi$ factor, both as a function of $\pir$ are shown in (c) and (d) respectively.
    }
    \label{fig:RX}
\end{figure}

The low-angle XRR data for samples with $\pir =$ 3 and 25.5 mTorr can be seen in Fig. \ref{fig:RX}(a). The periodicity of the oscillations (Kiessig fringes) is sensitive to the total thickness of the multilayer ($\tsamp$) \cite{Book-RX-Vaclav}.
The value of $\tsamp$ can be obtained fitting the modified Bragg's law \cite{Art-APL60-Nakamura, Art-APL112-Quinteros} 
\begin{equation}
    \sin{(\theta _{\mmin})} ^2 = 2 \delta + \mmin^2 \cdot \left ( \frac{ \lambda }{2 \tsamp} \right )^2,
    \label{Eq:RX}
\end{equation}
where $\lambda$ is x-ray wavelength, $1-\delta$ is the real part of the refractive index of the film and $\mmin$ the order of each local minima, $\theta_{\mmin}$ (see inset in Fig. \ref{fig:RX}(a)). From this technique a total value of the layer growth on the SiO$_2$ substrate of $\tsamp = 10 \pm 1$ nm was obtained.

In Fig. \ref{fig:RX}(b), a typical rocking curve measurement is shown for the sample grown with $\pir$ = 3 mTorr. This rocking curve was acquired around the first Kiessig fringe determined from the XRR measurements. From the rocking curves, the areas corresponding to diffuse scattering ($S_d$) and specular scattering ($S_s$) were calculated. The ratio $\rho = S_d/S_s$ is proportional to a global roughness of the multilayer \cite{Art-JAP69-Savage}. The $\rho$ values as a function of $\pir$ are shown in Fig. \ref{fig:RX}(c). From the figure, an increase in the roughness ($\rho$) of the interfaces as $\pir$ increases is evidenced. This trend may be attributed to the increase of the Ir grain size with increasing deposition pressures due to the lower kinetic energy of the Ir atoms arriving at the substrate \cite{Art-ThinSol558-Srinivas}.
In order to further quantify the interfacial quality of the Pt/Co/Ir structures, the XRR data was analyzed using GenX \cite{Art-JAC-Bjorck}, from which a roughness-intermixing parameter $\chi$ can be calculated.

In thin-film systems, roughness typically describes the height of variations in the structural morphology at an interface. In contrast, intermixing describes the thickness to which chemical species are alloyed at an interface. It is important to note that the $\chi$ parameter calculated from refinement of the XRR data does not distinguish between structural roughness and chemical intermixing between layers \cite{Art-PRB95-Wells}. Although a slight decrease in $\chi$ with increasing $\pir$ is apparent in Fig. \ref{fig:RX}(d), there is no substantial variation in the roughness-intermixing as a function of $\pir$. In concert with the rocking curve analysis presented in Fig. \ref{fig:RX}(c) (which demonstrated a clear increase in global roughness with increasing $\pir$), the unchanging nature of $\chi$ with $\pir$ might indicate that roughness is increased and intermixing is reduced when using higher $\pir$ values \cite{Art-PRB91-Lavrijsen}. However, further characterizations, such as cross-sectional transmission electron microscopy, are required to more precisely understand the nature of the interfaces in the Pt/Co/Ir structures.

\label{sec:results}
\subsubsection{Static magnetic characterization}

\begin{figure}[]
  \centering
    \includegraphics[scale=0.55]{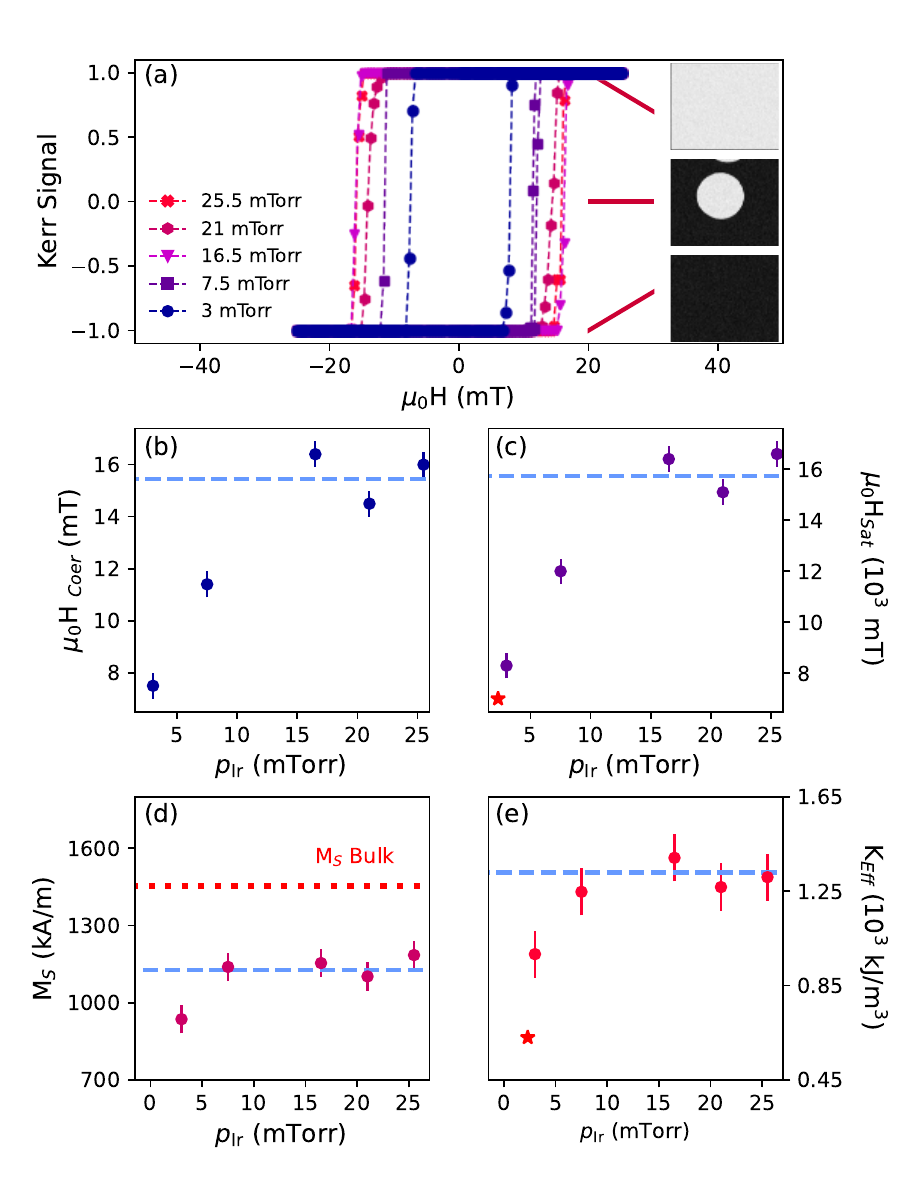}
    \caption{Magnetometry data for the Pt/Co/Ir structures acquired at room temperature for samples where the Ir layer was grown with different Ar pressures ($p_{\text Ir}$).
    (a) Magnetic hysteresis cycle. The insets show typical snapshots at the indicated fields. (b) Coercive field $\mu_0 \hcoer$. (c)  Saturation field $\mu_0 \hsat$.
    (d) Saturation magnetization $\ms$. (e) Effective anisotropy constant $\keff$.  Star symbols were taken from Ref.\cite{Art-PRB99-Shahbazi}. The red dashed line in (d) indicate the expected bulk value for $\ms$. The blue dashed lines indicate the values at which the different magnitudes saturate. Results shown in panel (a), (b) and (c) were measured by PMOKE microscopy, while the results shown in (d) and (e) using SQUID. 
    }
    \label{fig:Hist-SQUID-MOKE}
\end{figure}

We start by describing the $\pir$ dependence of the static magnetic properties as determined using PMOKE and SQUID via magnetization loops at room temperature with fields applied in-plane and out-of-plane, respectively. Fig. \ref{fig:Hist-SQUID-MOKE}(a) shows the loop obtained by PMOKE microscopy, where the Kerr signal is computed as the average value of the intensity of the whole set of pixels that form each image, that in our case corresponds to an areas of 0.187 mm$^2$. Typical snapshots are shown in the three insets for different fields, showing that the magnetization reversal is practically dominated by the nucleation and growth of domains. From these results, the coercive and saturation fields ($\hcoer$ and $\hsat$), shown in Fig. \ref{fig:Hist-SQUID-MOKE}(b) and \ref{fig:Hist-SQUID-MOKE}(c) , were obtained. 
Since the sample possesses strong perpendicular magnetic anisotropy (PMA), the values of $\hcoer$ and $\hsat$ differ by a maximum of 1 mT, indicating an almost square hysteresis loop. We observe an increase of $\hcoer$ and $\hsat$ with $\pir$ for low which tends to saturate at $\pir \approx 10$ mTorr.

The value of $\ms$ for a bulk Co system is 1440 kA/m, and all our samples present lower values. This phenomenology, generally observed in nanostructured systems and particularly in thin magnetic films similar to the one we are studying in this work, is usually associated with the presence of a magnetic dead layer~\cite{2023JMMM..58771339D, MDL_APL95_043106_2009, MDL_PRB.64.144422}. 
Similar trends for  $\ms$ and $\keff$ in Fig. \ref{fig:Hist-SQUID-MOKE}(d) and \ref{fig:Hist-SQUID-MOKE}(e) emphasize that both quantities are affected by $\pir$ in a similar way. 
A similar dependence with the Ar pressure within the deposition chamber was reported by Lavrijsen et al. \cite{Art-PRB91-Lavrijsen} for $\ms$, $\keff$ and also about the evolution of the shape of the hysteresis cycles in Pt/Co/Pt multilayers. Lavrijsen et al. associate this behavior to the sensitive nature of the perpendicular magnetic anisotropy (PMA) to the growth conditions. 
In a similar way Wells et al. \cite{Art-PRB95-Wells} have shown that, also for Pt/Co/Pt thin films, an increment of the Ar pressure within the deposition chamber induce a decrease of the intermixing between species of both layers, but at the same time an increment of the interface roughness.
Within the context of the structural characterizations presented in Figs.\ref{fig:RX}(c) and (d), the initial increase in $\hsat$, $\ms$, and $\keff$ with increasing $\pir$ followed by saturation at $\pir\approx 10$ mTorr observed is consistent with a chemically sharper, less intermixed Co/Ir interface at higher $\pir$. Conversely, the similar response of $\hcoer$ to $\pir$ is consistent with an increase in the structural roughness of the Co/Ir interface with $\pir$ \cite{Art-PRB95-Wells,Art-PRB91-Lavrijsen,Art-IEEE-Bandiera,Art-JMMM-Verbeno}.

\subsubsection{Domain wall velocity {under out-of-plane DC magnetic field}}
\label{sec:dwvel}

{We now analyze the DW average velocity at room temperature under DC magnetic fields. Within this section, all the magnetic fields, to nucleate and to grow the domains, were applied in the direction perpendicular to the plane of the sample, as described in   
}
{protocol (i) of Section \ref{sec:experiment}.}

\begin{figure}[]
  \centering
    \includegraphics[scale=0.5]{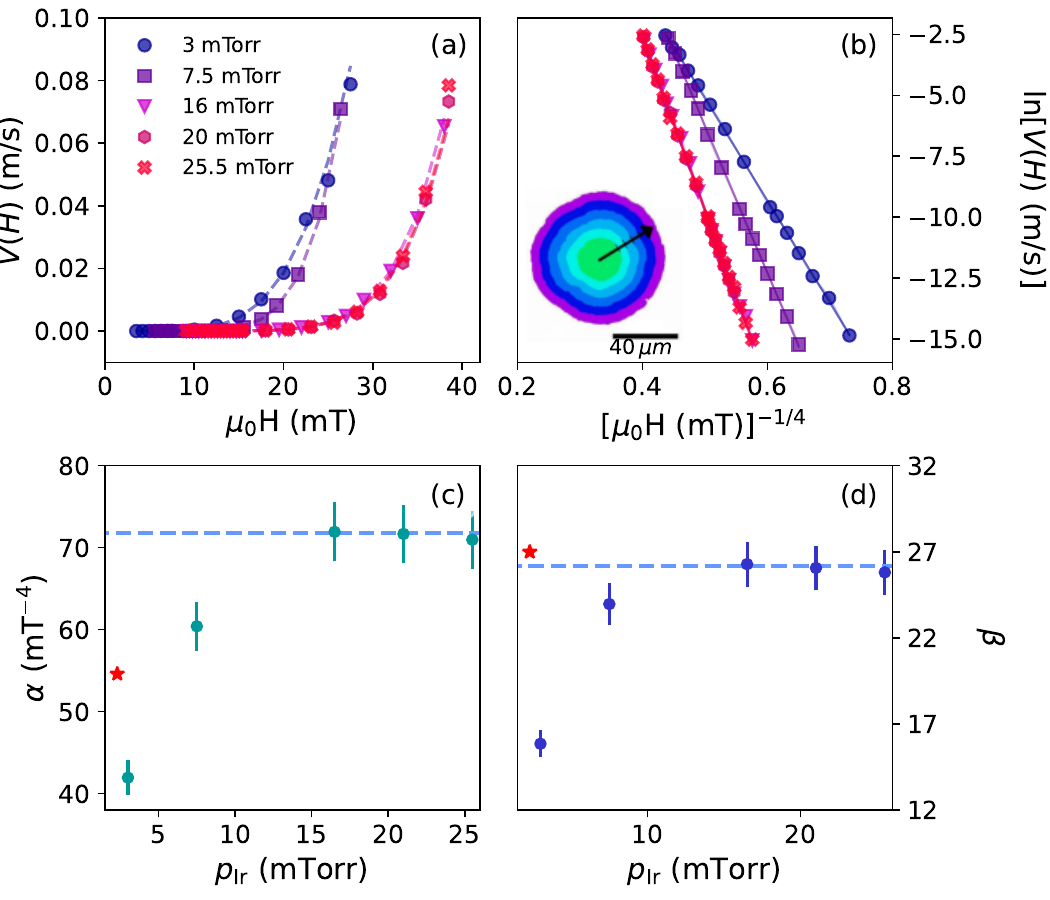}
    \caption{
    (a) Velocity field characteristics $V(H)$ for different Ir growth pressures $\pir$. 
    (b) Experimental measurements to the expression $\ln V(H) = -\alpha (\mu_0 H)^{-1/4} + \beta$, as expected from the creep-law  (Eq. (\ref{eq:VelCreep})), Solid lines are linear fits of the Eq. (\ref{eq:VelCreep}).
    Inset: Example of the growth of a domain obtained from the sample for $\pir =3$ mTorr, where several pulses of $\tau$=50 ms and $\mu _0 H=10$ mT are applied. A typical sequence of images of a domain growing under the action of the magnetic field pulses is shown in the inset. The images correspond to the sample with $\pir =3$ mTorr, under magnetic field pulses of $\tau =50$ ms and $\mu _0 H =10$ mT. From the linear fits we obtain $\alpha$ and $\beta$ vs. $\pir$, as shown in (b) and (c) respectively. 
    }
    \label{fig:Vel-MOKE}
\end{figure}

In Fig.\ref{fig:Vel-MOKE}(a) 
we plot $V(H)$ vs $H$ for different $\pir$, {as obtained from DW displacements such as the ones shown in the inset of Fig.\ref{fig:Vel-MOKE} (b)}. In 
Fig.\ref{fig:Vel-MOKE}(b), where we plot $\ln V(H)$ vs $(\mu_0 H)^{-1/4}$, we observe 
that the creep-law of Eq. (\ref{eq:VelCreep}) is fairly satisfied with $\mu=1/4$, as can be appreciated by the almost linear relationship between the two quantities.  
These fits allow us to obtain the $\pir$ dependencies of $\alpha$ and $\beta$, as shown in Fig.\ref{fig:Vel-MOKE}(c) and Fig.\ref{fig:Vel-MOKE}(d) respectively. 
For reference we also added the values of $\alpha$ and $\beta$ obtained in a similar stack~\cite{Art-PRB99-Shahbazi} for a single value of $\pir$.

From Figures \ref{fig:Vel-MOKE}(a)-(b) we observe that for a given field, $V(H)$ decreases with increasing $\pir$, remaining constant for $\pir \geq 16.5$ mTorr. This dependence is mainly controlled by the increase of $\alpha$ with $\pir$, specially at low $\pir$, as shown in Fig.\ref{fig:Vel-MOKE}(c).
Interestingly, not only $\alpha$ increases with $\pir$, but also $\beta$, becoming more important at higher fields. 
We also observe that for $\pir\approx 15$ mTorr both quantities tend to saturate.
The data taken from Ref.~\onlinecite{Art-PRB99-Shahbazi} is consistent with our results and specifically with the observed trend.
These $\pir$ dependencies evidence that interfacial, DC magnetic characteristics and DW dynamics, are correlated.

\subsubsection{Asymmetric domain growth}
\label{sec:asymmg}

The presence of DMI favours the formation of chiral DWs and produces an asymmetric domain growth in the presence of a perpendicular magnetic field ($H$) whenever a magnetic field in the direction parallel to the sample plane ($\hx$) is simultaneously applied. We will use this method to estimate the intensity of DMI, {applying the protocol (ii) of Section \ref{sec:experiment}.}

\begin{figure}[h]
  \centering
    \includegraphics[scale=0.65]{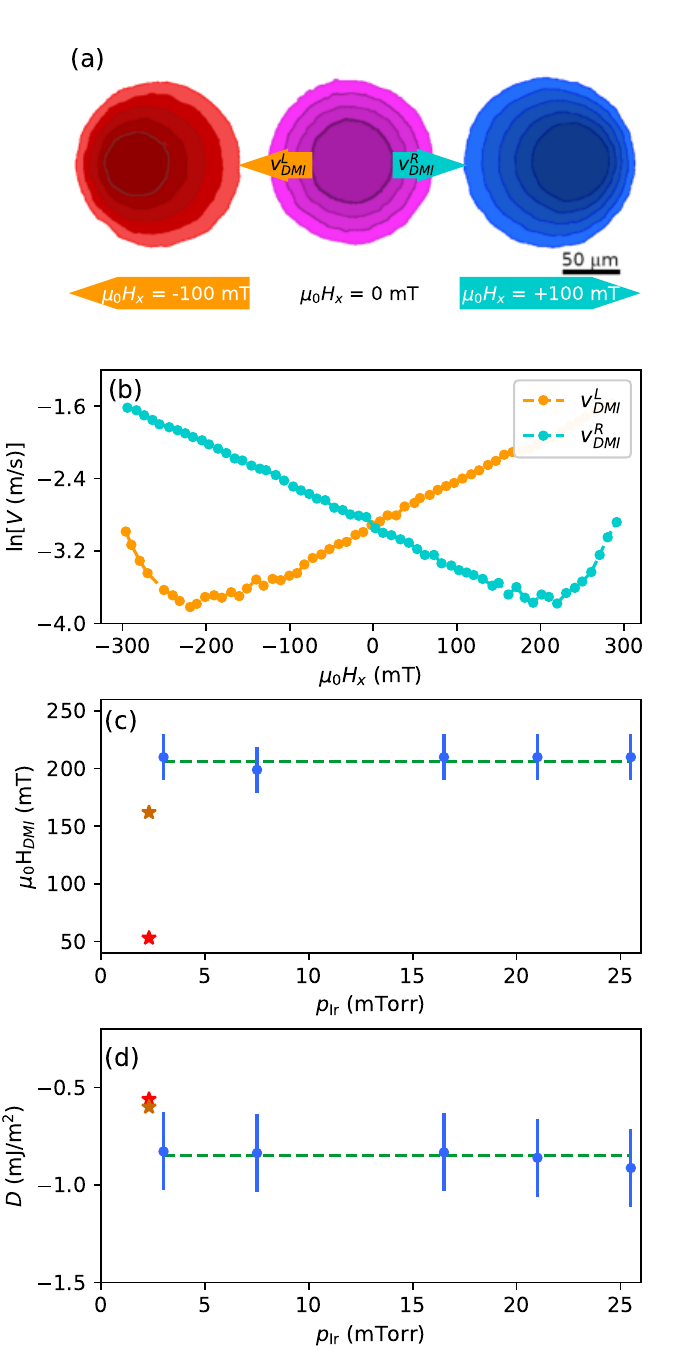}
    \caption{(a) Domain contours grown by applying perpendicular magnetic field pulses of an amplitude of $\mu_0 H = 20$ mT
    with $p _{Ir} = 3$ mTorr for a constant in-plane magnetic field $\mu_0 H_{\mathrm x} = \pm 100$ mT.
    (b) Typical velocities $v ^L _{\mathrm{DMI}}$ and $v ^R_{\mathrm{DMI}}$ measured for $\mu_0 H = 20$ mT pulses
    and $\mu_0 H _{\mathrm x} = \pm 100$ mT. (c) and (d): DMI field $\mu_0 \hdmi$ and 
    strength $D=\mu_0 \hdmi \ms \sqrt{A/\keff}$
    as a function of $\pir$, using a constant $A=17$ pJ/m. The dashed green lines indicate the average values, ${\mu_0 \hdmi} \approx 214$ mT, 
    ${D}  \approx -0.87$ mJ/m$^2$ 
    and the stars are values obtained in Ref.\cite{Art-PRB99-Shahbazi}, the red star from asymmetric growth and the orange from Brillouin light scattering (BLS).
    }
    \label{fig:Fig-VelPlan}
\end{figure}

Fig.~\ref{fig:Fig-VelPlan}(a) shows snapshots of the domains growing in the presence of a pulsed out-of-plane field in a constant in-plane field.
{
We observe that for $\hx=0$ the domain growth is symmetric 
while a clear left-right asymmetric growth is observed for $\hx \neq 0$, as indicated.}
In Fig \ref{fig:Fig-VelPlan}(b) 
we show the resulting right $\vrdmi$ and left $\vldmi$ velocities as a function of the in-plane field $\hx$ for a given $\pir=3$ mTorr. 
The curves display clear minima at $\mu_0\hdmi \sim \mu_0|\hx|\approx 214$ mT. In Fig. \ref{fig:Fig-VelPlan}(c) we show $\hdmi$ vs $\pir$ and also indicate the value obtained in Ref.\cite{Art-PRB99-Shahbazi} for a similar sample.

As can be appreciated from Fig. \ref{fig:Fig-VelPlan}(c), $\mu_0\hdmi\approx~214$ mT and there is no appreciable and systematic change with $\pir$ in the studied range. 
The fact that, for the entire range of $\pir$ of studied, we obtain an almost constant value for $\hdmi$ contrasts with the trend previously discussed for $\ms,\keff,\alpha,\beta$.
{
In Fig. \ref{fig:Fig-VelPlan}(d) we show 
the corresponding DMI strength $D$ from  Eq.\eqref{eq:DMIstrength},  
using 
$A=17$ pJ/m. We see that $D$ is practically constant for $\pir<15$ mTorr, around $D \approx -0.87 \text{ mJ/m}^2$ and for $\pir>15$ mTorr presents a slight increase in intensity. 
Both in Figures \ref{fig:Fig-VelPlan}(b)-(c) we compare with results from Ref.\cite{Art-PRB99-Shahbazi} for a particular value of $\pir$, obtained from asymmetric growth and Brillouin light scattering.}

\subsubsection{AC dynamics of domain walls}

The analysis of PMOKE images under alternating magnetic field pulses with zero mean value is similar to the one used in the DC case (Sec.\ref{sec:dwvel}) 
except that in this case, after nucleation of the initial domain, pictures are taken after applying one positive square pulse followed by one negative square pulse of the same amplitude $H$, {as described by protocol (iii) in Section \ref{sec:experiment}}
(from now one we refer to this pair of square pulses as the single ``AC pulse''). Due mainly to the finite (negative) mean curvature of the domain and the presence of quenched disorder, the dynamics is expected to be irreversible, yielding complementary information to the DC analysis. Such information can be partially captured by monitoring simple geometrical properties such as the area, the perimeter and the roughness of the closed domain wall~\cite{Art-PRB103-Domenichini}. 

\begin{figure}[]
  \centering
    \includegraphics[scale=0.7]{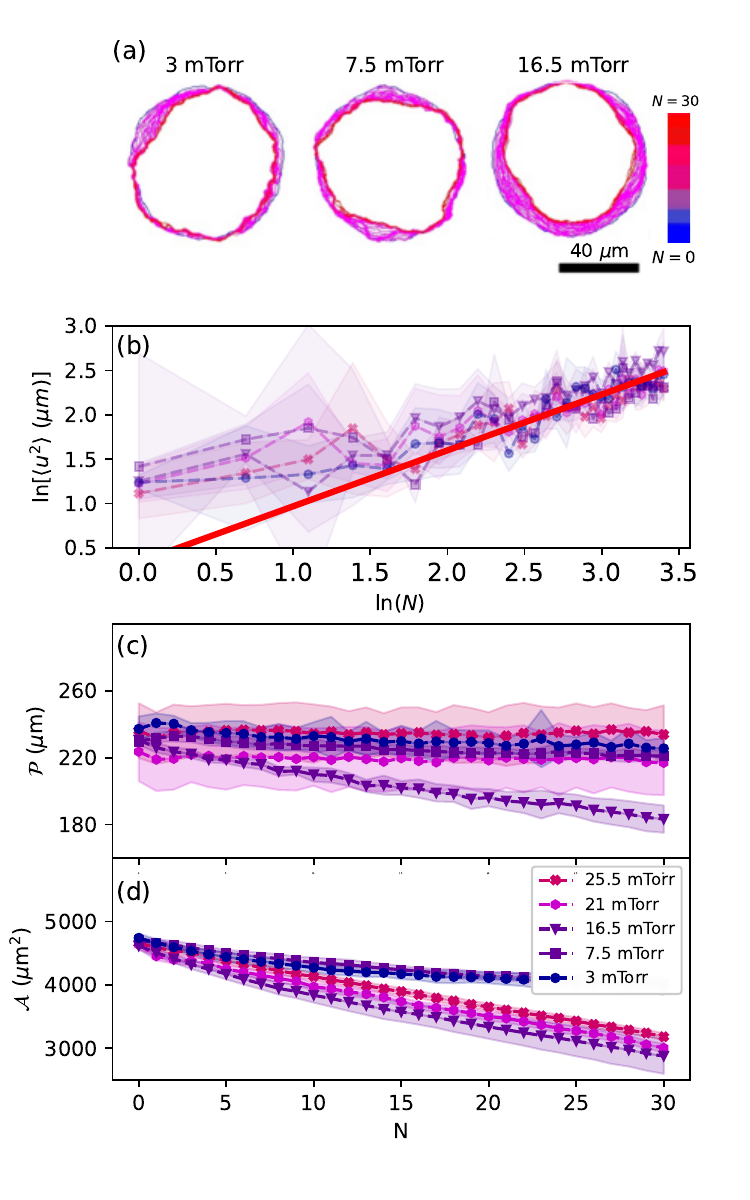}
    \caption{
    (a) Contours images showing the DW shrinking and dynamical roughening as a function of the AC pulse number $N$, from $N=0$ (blue contour) to $N=30$ (red contour). Samples are grown at the different $\pir$ as indicated. (b) Mean square displacement $\langle u ^2 \rangle$ vs $N$ for four samples grown at different $\pir$. The red line describes $\langle u ^2 \rangle \sim N^{\gamma}$, with $\gamma = (0.63 \pm 0.01)$. (c)-(d) Evolution of the perimeter $\perimeter$ and area $\area$ of the domain as a function of $N$ for the four samples.
    } 
\label{fig:Fig-PerAr}
\end{figure}

Fig. \ref{fig:Fig-PerAr}(a) shows domain walls snapshots as successive AC pulses of magnetic field were applied for three samples grown at different $\pir$.
The colors of the contours indicate the number of pulses applied before taking the picture, starting with blue contours ($N = 0$) and ending with red contours ($N = 30$). The domain evolution as a function of pulse number $N$ is characterized by an increase of the domain wall roughness concomitantly with the decrease of the domain area and perimeter, expected from curvature effects.

\begin{figure}[]
  \centering
    \includegraphics[scale=0.65]{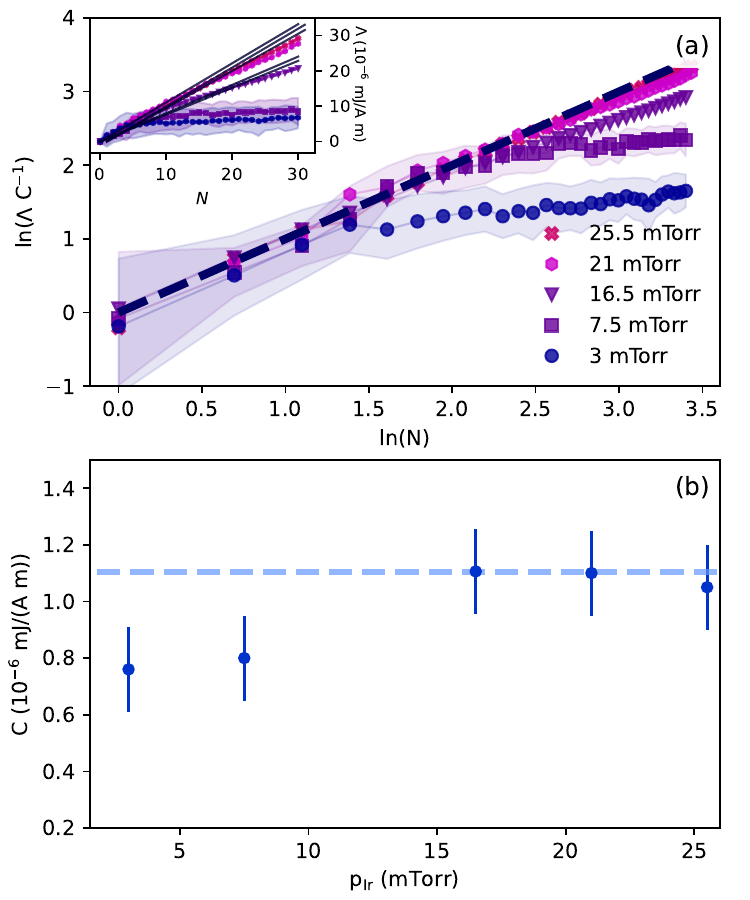}
    \caption{(a) Rescaled function $\Delta \Lambda = (\Lambda(N)-\Lambda(0))$ with $C$ a $\pir$-dependent quantity. (b) Values of $C$ obtained from the adjustments of $\Lambda$ as a function of $\pir$, where the dashed-line is the average value of $C$ for all samples. 
    }
    \label{fig:Fig-Lambda}
\end{figure}

The roughening of the domain wall is shown in Fig. \ref{fig:Fig-PerAr}(b). We clearly observe two regimes for the mean square displacement $\langle u ^2 \rangle$ as a function of the AC pulse number $N$, as obtained from the domain wall contours (see Sec.\ref{sec:properties}). 
For the five samples in Fig. \ref{fig:Fig-PerAr}(b) we observe, that $\langle u ^2 \rangle \sim N^{\gamma}$ with $\gamma = (0.63 \pm 0.01)$, consistent with a typical dynamic roughening process~\cite{barabasi_stanley_1995}.
In Fig. \ref{fig:Fig-PerAr}(c) and (d) we show the evolution of the perimeter $\perimeter$ and the area $\area$ respectively as a function of $N$. We observe a decrease in $\area$, as seen in previous works \cite{Art-PRB99-Domenichini, Art-PRB103-Domenichini}. 
The decrease in $\perimeter$ on the other hand is determined by the competition between the increasing roughness, which tends to increase it and the decreasing area which tends to reduce it.
We observe that both $\area$ and $\perimeter$ decrease faster at larger $\pir$ and tend to saturate on the same decay for lower pressures.
Using the data of Fig. \ref{fig:Fig-PerAr}(c)-(d), and the DC velocity characteristics of Fig. \ref{fig:Vel-MOKE} we can now compute the function $\Lambda (N)$ using Eq.(\ref{eq:Lambda}). 
In Fig. \ref{fig:Fig-Lambda}(a) we show $\Delta \Lambda(N) \equiv \Lambda(N)-\Lambda(0)$ for all samples with different $\pir$. For each one we can fit a linear function whose slopes $C$ are a function of $\pir$.
From the plot of ln$(\Delta \Lambda \cdot C^{-1}$) vs. ln$(N)$ shown in the main panel, we can see that the approximation of Eq.(\ref{eq:Lambda}) ($\Delta \Lambda(N) \approx C N$) is valid for all $\pir$ up to $N \approx 10$.
In Fig. \ref{fig:Fig-Lambda}(b) we show that the fitted values of $C$ grow with $\pir$ with a similar behaviour as $\hcoer$, $\hsat$, $\ms$ and $\keff$.

\section{Discussion} \label{Sec:Disc}
\label{sec:discussion}

Our results show that the micromagnetic parameters $M_s$, $\keff$, $\hcoer$, $C$, share a similar dependence with $\pir$. We can see that all quantities increase with increasing $\pir$ and saturate at $\pir \approx 15$ mTorr. In all cases, the saturation value is roughly a factor 1.5 or 2 larger than the value of the same quantity at the lowest pressure $\pir=3$ mTorr.

A similar dependence but as a function of the growth pressure of the Pt top layer was observed in Pt/Co/Pt \cite{Art-PRB91-Lavrijsen}. On that case the authors attributed the observed trend to a possible crossover between two different regimes.  On the one hand, a regime of considerable alloying or intermixing between Pt and Co at low pressures and, on the other hand, a regime of layered growth of the Pt layer above a given crossover pressure.

\subsubsection{DC creep dynamics}
Regarding the DC dynamics properties it is interesting 
to analyze first the dependence with $\pir$ of the creep parameters $\alpha$ and $\beta$, defined in Eq.(\ref{eq:VelCreep}).
The values of these parameters were
fitted to the results obtained for the velocity as a function of the field in the creep regime. As shown in Fig.\ref{fig:Vel-MOKE}(b)-(c) these two parameters display a very similar dependence with $\pir$, with a saturation at $\pir\approx 15$ mTorr reaching a value which is roughly a factor 1.5 or 2 larger than the value of the same quantity at the lowest pressure $\pir=3$ mTorr. Note that a very similar trend is observed for the micromagnetic quantities shown in Fig.\ref{fig:Hist-SQUID-MOKE}(b)-(e).

This behaviour can be rationalized by combining the extended creep law \cite{jeudy2016} and weak pinning theory. The extended creep law is the empirical extrapolation of the low field velocity, $v \propto e^{-(T_d/T)(\hd/H)^{1/4}}$ to larger fields, even up to $\hd$. The extrapolation reads 
$v \propto e^{-U(H)/k_B T}$ with   
the extended universal creep barrier $U \approx k_B T_d [(\hd/H)^{1/4}-1]$, $k_B$ the Boltzmann constant, $T_d$ the characteristic temperature scale and $\hd$ the depinning field.
The extended barrier $U(H)$ hence accounts for both, the divergence in the $H\ll \hd$ limit, and its expected vanishing by approaching $\hd$. 
This extended though empiric form was shown to hold for a large family of thin-film materials up to $\hd$ \cite{jeudy2016}. We can thus write
\begin{eqnarray}
    v \approx \vd e^{-(T_d/T)[(\hd/H)^{\mu}-1]}. 
\end{eqnarray}
from where we can identify, using Eq.(\ref{eq:VelCreep}),
\begin{eqnarray}
\alpha &=& \hd^{\mu} T_d/T,
\label{eq:alpha}
\\
\beta &=& (T_d/T) +\ln (\vd),
\label{eq:beta}
\end{eqnarray}
In Eq(\ref{eq:beta}) we are fixing the units of velocity to m/s.
Using weak collective pinning theory \cite{Art-PRB98-Jeudy} we get 
\begin{eqnarray}
\hd &\approx& \frac{\sigma_0 \xi}{2M_{\mathrm{S}} L_c^2}, 
\\ 
T_d &\approx& \frac{\sigma_0 \xi^2 t}{L_c},
\end{eqnarray}
where $L_c$ is the Larkin length~\cite{blatter_vortex_review,gorchon2014},
\begin{equation}
    L_c \approx 
    \left(\frac{\sigma_0^2 \xi^2}{t^2 \Delta_0}\right)^{1/3},
\end{equation}
where $t$ is the average magnetic film thickness,
$\sigma_0 \approx 4\sqrt{A \keff}$ the DW energy per unit area, $\Delta_0$ the strength of the pinning forces and $\xi = \max[\delta, \xi_0]$ the correlation length of the pinning force acting on the DW. Here, $\xi_0$ is a constant determined by the random spatial heterogeneities.
Therefore,
\begin{eqnarray}
\hd &\approx& \left(\frac{\Delta_0^{2} t^{4}}{8 \ms \sigma_0 \xi} 
\right)^{1/3}, 
\label{eq:hdmicro}
\\ 
T_d &\approx& (\Delta_0 \sigma_0 t^{5} \xi^{4})^{1/3},
\label{eq:Tdmicro}
\end{eqnarray}
implying that both $\alpha$ and $\beta$ are proportional to 
\begin{eqnarray}
\alpha, \beta \propto
\Delta_0^{1/3} t^{5/3}(\sigma_0 \xi^4)^{1/3}.
\label{eq:alphabetamicro}
\end{eqnarray}
To estimate $\alpha$, we have neglected the weak contribution of $\hd$ in comparison to $\td$, as all the micromagnetic parameters of $\hd$ in Eq.(\ref{eq:hdmicro}) are raised to the small power $1/12$ or smaller. We have also neglected the contribution $\ln(\vd)$ in $\beta$ because for most known samples displaying a universal creep regime we have $\td \gg T$ and from thermal rounding experiments we know that $\vd \equiv v(H=\hd,T)$ is typically of the order of $1-10$ m/s \cite{jeudy2016}. 
Under these assumptions a similar dependence with $\pir$ is expected for both $\alpha$ and $\beta$. 
This estimate agrees with the results of Fig.\ref{fig:Vel-MOKE}(b)-(c).

\begin{figure}[h!]
  \centering
    \includegraphics[scale=0.55]{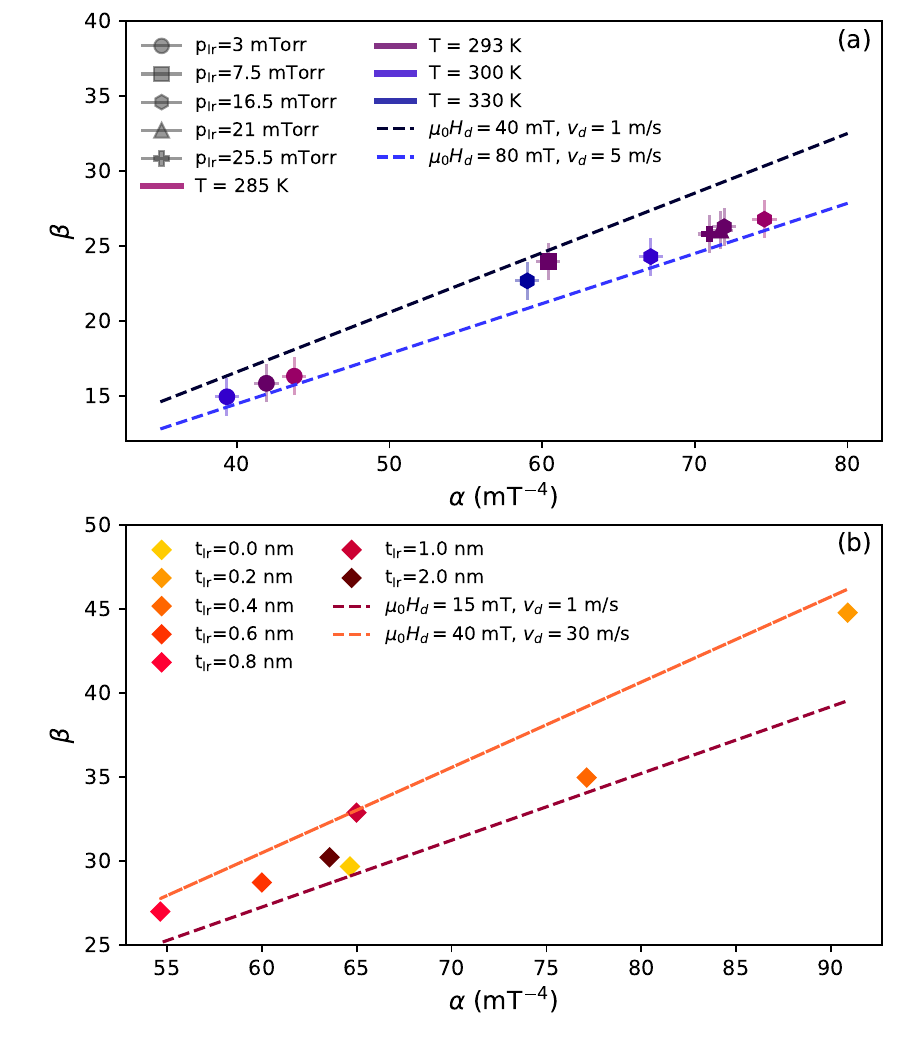}
    \caption{Creep-law parameters $\beta$ vs $\alpha$ for different Ta/Pt/Co/Ir/Ta samples. (a) Samples with different $\pir$ (plot style). and temperatures (colour). (b) Samples of Ref. \cite{Art-PRB99-Shahbazi} with different Ir thickness ($t_{\mathrm{Ir}}$). Dotted lines indicate the bounds expected from 
    Eq.(\ref{eq:correlationalphabeta}) 
    using the known range of $\hd$ and $\vd$.
    }
    \label{fig:Beta-Alpha}
\end{figure}

From Eqs.(\ref{eq:alpha})-(\ref{eq:beta}) we can readily write 
\begin{equation}
    \beta = \hd^{-1/4} \alpha + \ln(v _d), \label{eq:correlationalphabeta}
\end{equation}
implying a temperature independent linear correlation between $\alpha$ and $\beta$, as long as $\hd^{-1/4}$ and $\ln \vd$ does not vary appreciably for the range of $\alpha$ and $\beta$ analyzed.  
In Fig.\ref{fig:Beta-Alpha}(a) we show that a similar linear correlation holds for $\alpha$ and $\beta$ obtained at different $\pir$ in our samples and also for those pairs of values $\alpha$ and $\beta$ obtained by Shahbazi \textit{et al.} \cite{Art-PRB99-Shahbazi} but for different Ir thicknesses. We observe that although the films have the same composition and the slopes (related to $\hd^\mu \td$) are quite similar, the intercept (related to $\vd$) differ appreciably. Eqs. (\ref{eq:correlationalphabeta}) explain all these observations and provide an approximated method to obtain $\hd$ and $\vd$. 
In particular, the linear fit allows us to obtain $\hd = 74 \pm 6$ mT and $\vd = (10 \pm 1)$ m/s for our whole set of samples with different $\pir$. These values are in very good agreement 
with the estimate of $\hd$ and $\vd$ from velocity curves measured from low to high fields $H>\hd$,  displaying deviations from the creep-law \cite{Art-PRB99-Shahbazi}. 
Interestingly, such linear correlation has been also observed in several different ultrathin ferromagnetic stacks ~\cite{Art-APL112-Quinteros}. In the appendix we show that the predicted linear correlation of Eq.(\ref{eq:correlationalphabeta}) allows to obtain consistent bounds for $\hd$ and $\vd$ for these cases. In summary, these correlations between $\alpha$ and $\beta$ are naturally explained in terms of the universal extended creep law~\cite{jeudy2016}.

{In order to obtain a more microscopic insight, we analyze Eq. (\ref{eq:alphabetamicro}). Assuming that disorder (through $\Delta_0$) does not depend on $\pir$, two situations may be considered.
\begin{enumerate}
\item[(i)] On the one hand, if $\xi = \xi_0>\delta$ and that the exchange parameter $A$ is independent of $\pir$, we get $\alpha,\beta \propto \keff^{1/6}$, which is consistent with the results of Fig.\ref{fig:Hist-SQUID-MOKE}(e) for $\keff$ and Fig.\ref{fig:Vel-MOKE}(b) and (c) for $\alpha$ and $\beta$. That is, $\alpha$, $\beta$, and $\keff$ increase with increasing $\pir$ and saturate at $\pir\approx 15$ mTorr.
\item [(ii)] On the other hand, if there is point pinning, then $\delta > \xi_0 \Rightarrow \xi = \delta \approx \sqrt{A/\keff}$.  In this case we get $\alpha,\beta \propto \keff^{-1/2}$ 
which is apparently inconsistent with the $\pir$ dependencies of the three quantities.
In order to reconcile in this case the $\pir$ dependencies of Fig.\ref{fig:Hist-SQUID-MOKE}(e) for $\keff$ and Fig.\ref{fig:Vel-MOKE}(b)-(c) 
we need to assume that $A^{3/2}$ grows with $\pir$ faster than $\keff^{1/2}$, so that their product $A\keff$ grows with $\pir$. 
\end{enumerate}
}

{
Deciding between (i)-(ii) strictly requires a study of $A$ vs $\pir$. Nonetheless, it is worth noting that scenario (i) appears to be consistent, as $\xi$ can be also estimated from Larkin weak collective pinning theory \cite{Art-PRB98-Jeudy,Gehanne2020} and satisfies $\xi>\delta$. The corresponding expression is  
\begin{equation}
    \xi \approx \left[
    \frac{(k_B T_d)^2}{2\mu_0 H_d M_s \sigma_0 t^2} 
    \right]^{1/3},
\end{equation}
and all data for the estimate is available:
From Figure \ref{fig:Beta-Alpha}(a) we get that $40 \text{ mT } < \mu_0 \hd < 80 \text{ mT }$; From $\beta$ in Table \ref{Tabla:Constantes} and equation \eqref{eq:beta} (neglecting again the small contribution of $\ln(v_d)$), we get $4000 \text{ K } < T_d < 7500 \text{ K }$ for the whole $\pir$ range; From Table \ref{Tabla:Constantes} we get $\ms$ and $\sigma_0 = 4\sqrt{A K_{\rm eff}}$ for the whole $\pir$ range; By fixing values $A=17 \text{ pJ/m}$ and $t \approx 0.8$ nm we get values of $\xi$ which can be directly compared to the DW-width estimate $\delta \approx \sqrt{A/\keff}$ vs $\pir$, as shown in Table \ref{Tabla:Constantes}. As can be appreciated, $\xi \gtrsim 3 \delta$ for the whole $\pir$ range. One should nevertheless be cautious as this result combines several estimates, both from pinning theory and from the micromagnetic theory. 
}

\subsubsection{AC creep dynamics}

\begin{figure}[]
  \centering
    \includegraphics[scale=0.6]{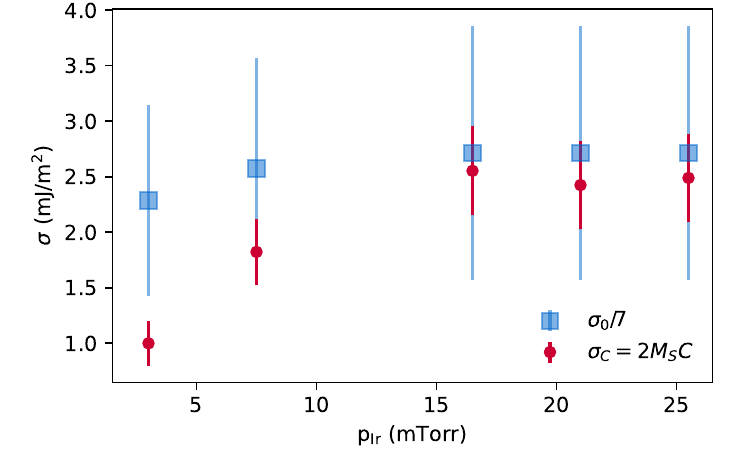}
    \caption{
    Phenomenological domain wall tension $\sigma_C$ obtained from the effective area $\Lambda (N)$ obtained in AC measurements,  compared to the ratio $(\sigma_0/7)$ vs $\pir$, with $\sigma_0=4\sqrt{A \keff}$ the micromagnetic estimate, as a function of $\pir$.
    }
    \label{fig:SigmaCy0}
\end{figure}

We now show how the AC field creep response can yield supplementary information to the DC field creep response.
In Fig.\ref{fig:Fig-Lambda} we show that the effective area can also provide insights into the effects of changing $\pir$. The evolution of the geometrical observable $\Lambda$, defined in Eq. (\ref{eq:Lambda}) displays two regimes as a function of the number of AC cycles $N$, as can be appreciated in Fig.\ref{fig:Fig-Lambda}(a). For small $N$, $\Lambda(N)-\Lambda(0) \propto C N$ as predicted by Eq.(\ref{eq:Lambda}), allowing us to estimate the parameter $C$, shown in Fig.\ref{fig:Fig-Lambda}(b) as a function of pressure $\pir$. Interestingly, it displays the same trend than $\ms,\keff,\hcoer,\alpha,\beta$, i.e. increasing from low $\pir$ and saturating at $\pir \approx 15$mTorr (the values of all these parameters are available in Table \ref{Tabla:Constantes}). As explained in Ref.\cite{Art-PRB103-Domenichini}, $C \sim \sigma_0/2\ms$ provided certain conditions are met. In particular, an homogeneous creep law response of the DW at its different points is assumed to derive Eq. (\ref{eq:Lambda}), and this approximation was shown to be better satisfied for low values of $N$, both in simulations and experiments~\cite{Art-PRB103-Domenichini}.
In Fig.\ref{fig:SigmaCy0} we show that the estimated elastic tension, $\sigma_C = 2 \ms C$, using the measured values of $C$ (Fig.\ref{fig:Fig-Lambda}(b)) and $\ms$ (Fig. \ref{fig:Hist-SQUID-MOKE}(d)), changes with $\pir$ in a very similar fashion than $\sigma_0=4\sqrt{A \keff}$ does, using $A=17$~pJ/m \cite{Art-PRB99-Shahbazi} and the measured $\keff$ of Fig.\ref{fig:Hist-SQUID-MOKE}(e). Nevertheless, we also find a whole factor $\sim 7$ of difference which remains to be fully explained. 
One possibility is that DMI is correcting $\sigma$, as suggested by Je {\it et al.} \cite{Art-PRB88-Je}. The predicted correction for our case is however insufficient because it is of order $\sim 10$ $\%$. 
Another possibility can be related to thickness modulations. In this case, as speculated in relation to the $\pir$ dependencies of $\alpha$ and $\beta$, a possible explanation is that the effective average thickness $t_{\mathrm{eff}}$ seen by the dominant metastable states is systematically smaller than the magnetic thickness $t_{\mathrm{M}}$ because they profit thinner regions to minimize the local DW energy. This would correct $\sigma_0$ effectively by 
$\sigma_{\mathrm{eff}}\approx \sigma_0 t_{\tt eff} /t_{\mathrm{M}} < \sigma_0$, changing $\sigma_{\mathrm{eff}}$ in the direction of the phenomenological $\sigma_C$. Additionally, $\sigma_C$ may incorporate other inhomogeneity effects which are absent in the micromagnetic bulk $\sigma_0$.
Besides the discussed difference, however, the predicted $\pir$ dependence from the AC theory seems to be consistent with the changes of the micromagnetic parameters with $\pir$. The effect of $\pir$ on $\ms,\hcoer,\keff,\sigma$ is in sharp contrast with its effect on $\hdmi$. 

The AC field protocol can also yield information about the pinning potential for larger number of AC pulses $N$, through the analysis of the AC field induced roughening of the domain walls characterized in Figs. \ref{fig:Fig-PerAr}(a)-(b). 
We find that for for $N \gtrsim 7$ the average quadratic width of the wall grows as $\langle u^2 \rangle \approx B N^{\gamma}$, with $\gamma \approx 0.63\pm 0.01$, and $B$ a constant prefactor. 
Consistently, $N \sim 7$ matches the onset of deviations from the the predicted scaling $\Lambda(N)-\Lambda(0)\approx C N$ (see Eq. (\ref{eq:Lambda})), as shown in Fig. \ref{fig:Fig-Lambda}(a). This is somehow expected as such linear behaviour is predicted by assuming that the domain wall local response is homogeneous, an assumption that it is expected to hold only for the first few pulses, when the nucleated DW is still largely uncorrelated with the underlying pinning potential. Roughening indicates, consistently, that the response becomes heterogeneous and that the domain wall configuration gets  more correlated with the underlying pinning potential as $N$ grows. These correlations add relevant information through the exponent $\gamma$ and the prefactor $B$. If we assume that the large-scale geometry of a DW in the AC creep regime is described by the Edwards-Wilkinson (EW)
equation with an effective temperature, as it occurs in the DC creep case
~\cite{chauve2000,kolton2009creep,grassi2018}, we expect a roughening process characterized by the dynamical exponent $z=2$ and a roughening exponent $\zeta=1/2$, so $w_t^2 \sim t^{1/2} \sim N^{1/2}$~\cite{barabasi_stanley_1995}, with the effective time dictated by the pulse number $N$ and the pulse duration $\tau$ as $t \sim N\tau$. This prediction clearly underestimates the result of~\ref{fig:Fig-PerAr}(b) for $\gamma$. A possible explanation for the difference can be attributed to the fact that under an AC field the effective time dependent noise acting on the domain wall is more correlated in time than the effective noise on the DC case, because in the former the DW can revisit the same disorder in its oscillatory motion. This makes the effective large-scale noise time correlated or colored. If we model such colored noise $\eta(x,t)$
with an exponent $\psi>0$, such that 
for two points in a DW segment
$\langle \eta(x,t)\eta(x,t')\rangle \sim \delta(x-x')|t-t'|^{2\psi-1}$ ($\psi=0$ for uncorrelated noise)
linear theory predicts $z=2$ and $\zeta=1/2+2\psi > 1/2$~\cite{barabasi_stanley_1995}. 
Therefore $w^2_t$ or $w^2_N$ should grow {\it faster} than in the DC case, with $\gamma=1/2+2\psi>1/2$. Our data agrees with this prediction with $\psi=0.065$ indicating a weak non-zero time correlation of the effective noise. The prefactor $B\approx \langle u^2 \rangle/N^{\gamma}$ on the other hand, is expected to quantify the strength of the disorder or effective noise. As can be appreciated in Fig. \ref{fig:Fig-PerAr}(b) 
the change is hardly detected for different growth pressures, thus indicating that growth pressure $\pir$ does not have an important impact in the sample disorder. It would be interesting to test this roughening phenomenon with numerical DW simulations under AC field in random media.

\section{Conclusions}
\label{sec:conclusion}

Summarizing, we have characterized the static and DC/AC dynamic properties of DWs in perpendicularly magnetized Ta/Pt/Co/Ir ultrathin films as a function of the Ir layer growth pressure.
We find that the saturation magnetization, the effective anisotropy constant and the domain wall surface tension increase monotonically with the growing pressure of the Ir layer and saturate at a threshold pressure. In sharp contrast the Dzyaloshinskii-Moriya field and the strength of the disorder remain almost constant across the same pressure range. 

\begin{acknowledgements}

We express our gratitude to Anni Cao, Avinash Chaurasiya, Katherine Nygren, and Pierre Vallobra for their work on the project ``A Comparison of Dzyaloshinskii-Moriya Interaction Measurement Techniques in Pt/Co/Ir Thin Films'', funded by the IEEE Magnetic Society. Additionally, we extend our thanks to Eric Fullerton for his valuable contribution to the growth of the samples.
We wish to thank Luis Avilés-Félix for the careful reading and constructive criticism of the manuscript. We thank Mara Granada for illuminating discussions.
This work was partially supported by the National Scientific and Technical Research Council - Argentina (CONICET), the University of Cuyo by grants C014-T1 and 06/C035-T1  and the ANPCyT by grants PICT2019-1991 and PICT 2019-2873.
\end{acknowledgements}

\appendix

\section{ \label{AppCreep} Creep analysis}

\begin{figure}[]
  \centering
    \includegraphics[scale=0.6]{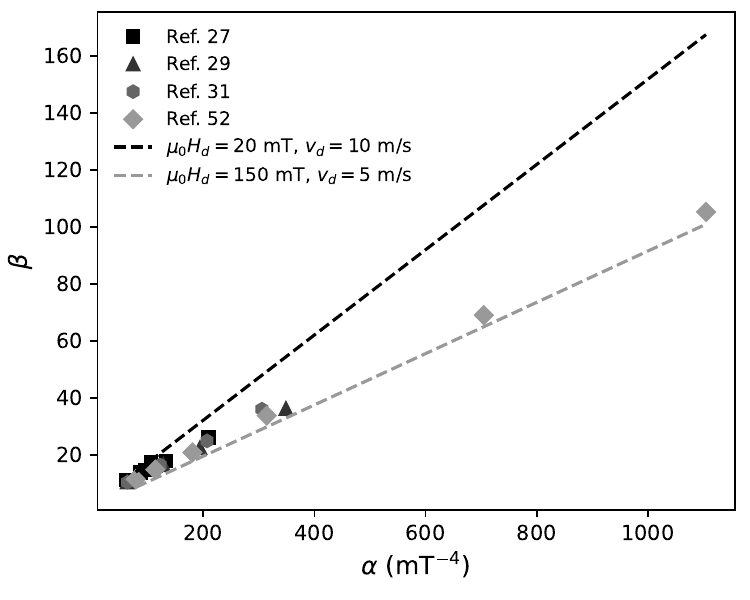}
    \caption{Interdependence of the creep law parameters $\alpha$ and $\beta$ obtained by Quinteros \textit{et al.} \cite{Art-APL112-Quinteros}, using experimental data from references \cite{gorchon2014,Art-PRL99-Metaxas,Art-PRB95-Pardo}. Using Eq. (\ref{eq:correlationalphabeta}) we obtain values of $\mu_0 \hd$ between 20 and 150 mT, and $\vd$ between 5 and 10 m/s.
    }
    \label{fig:DatosQuintero}
\end{figure}

In Ref. \cite{Art-APL112-Quinteros}, Quinteros \textit{et~al.}  analyzed the relationship between $\alpha$ and $\beta$, concluding that there is a linear correlation between them, as can be appreciated in Fig. \ref{fig:DatosQuintero}. We can interpret this relationship from Eq. (\ref{eq:correlationalphabeta}), which implies that the slope $d\beta/d\alpha$ depends on $\hd$ and the intercept on $\vd$.

From references \cite{gorchon2014,Art-PRL99-Metaxas,Art-PRB95-Pardo} we obtain values of $\mu_0 \hd$ between 20 and 150 mT and $\vd$ between 10 and 5 m/s. As we can see in Fig. \ref{fig:DatosQuintero}, the points are bounded by Eq. (\ref{eq:correlationalphabeta}) with values, corresponding to ($\hd=20$ mT, $\vd= 10$ m/s) and ($\hd=150$ mT, $\vd=5$ m/s).
Note that the linear relation observed for high $\alpha$ and $\beta$ values is different from the one observed at small values. This fact highlights a linear but non universal relation between both quantities. 
In this way, instead of defining a single value for all the samples, we define a ranges of $\hd$ and $\vd$ that correspond to the measurements of these parameters reported in all the references. 
Eq. (\ref{eq:correlationalphabeta}) hence explains the apparent correlation between $\alpha$ and $\beta$ across different samples.

\section{ \label{Tabla} Sample parameters}

In Table \ref{Tabla:Constantes} we show the values of the different constants obtained by the static and dynamic characterizations. $\ms$, $\heff$, were obtained from the hysteresis loops, $\keff$ was calculated as $\mu_0\heff \ms^2/2$, $\alpha$ and $\beta$ was obtained from the lineal fits from the creep-law of $V(H)$ vs $H^{-1/4}$, $\mu_0\hdmi$ was obtained from the asymmetric domain growth, $\sigma_0 =4\sqrt{A\keff}$ with $A = 17$ pJ/m, $C$ was obtained from effective area of AC dynamic, and $\sigma_C = 2 \, \ms \, C$. It is observed that the different properties of the samples increase with $\pir$, taking a saturation value for $\pir > 8$mTorr. These values are similar to those reported in some works in the bibliography \cite{Art-PRB99-Shahbazi, Art-PRB97-Shepley}. Furthermore, although they are not the same samples, the trends correspond to what was observed by Lavrijsen \textit{et al.} \cite{Art-PRB91-Lavrijsen}.

\begin{table*}[]
\begin{tabular}{|l|c|c|c|c|c|}
\hline
$\pir$ {(mTorr)}                            & 3           & 7.5         & 16.5        & 21          & 25.5        \\ \hline
$\ms$ {(kA/m)}                              & 940 (60)    & 1140 (60)   & 1150 (60)   & 1100 (60)   & 1190 (60)   \\ \hline
$\mu _0 \heff$ {(T)}                       & 2.1 (0.1)    & 2.2 (0.1)    & 2.4 (0.1)    & 2.3 (0.1)    &  2.2 (0.1)    \\ \hline
$\keff$ {(kJ/m$^3$)}                        & 1000 (100)  & 1200 (100)  & 1400 (100)  & 1300 (100)  & 1300 (100)  \\ \hline
$\delta$ (nm)     & 4.0 (0.5)      & 3.6 (0.5)      & 3.4 (0.5)      & 3.6 (0.5)      & 3.5 (0.5)      \\ \hline
$\xi$ (nm)     & 14 (7)      
& 13 (6)      & 12 (6)      
& 13 (6)      & 12 (6)      \\ \hline
$\alpha$ {(}[mT]$^{-4}${)}     & 42 (1)      & 60 (1)      & 72 (1)      & 72 (1)      & 71 (1)      \\ \hline
$\beta$                                       & 15.8 (0.1)      & 24.0 (0.1)      & 26.3 (0.1)      & 26.0 (0.1)      & 25.8 (0.1)      \\ \hline
$\mu _0 \hdmi$ {(mT)}                    & 210 (20)    & 200 (20)    & 210 (20)    & 210 (20)    & 210 (20)    \\ \hline
$\sigma _0$ {($10 ^{-3}$ J/m$^2$)}     & 16 (6)  & 18 (7)  & 19 (8)  & 19 (8)  & 19 (8)  \\ \hline
C {($ 10 ^{-6}$ mJ/Am)}              & 0.5 (0.1)   & 0.8 (0.1)   & 1.1 (0.1)   & 1.1 (0.1)   & 1.0 (0.1)   \\ \hline
$\sigma _C$ {($10 ^{-3}$ J/m$^2$)}     & 0.9 (0.2)        & 1.8 (0.3)       & 2.5 (0.4)        & 2.4 (0.4)        & 2.4 (0.4)        \\ \hline
$D$ {($10 ^{-3}$ J/m$^2$)}     & -0.8 (0.2)        & -0.8 (0.2)       & -0.8 (0.2)        & -0.9 (0.2)        & -0.9 (0.2)        \\ \hline
\end{tabular}
\caption{
Parameters of the full set of samples obtained through the static and dynamic characterization described in the main text.
}
\label{Tabla:Constantes}
\end{table*}

\bibliography{ref}

\end{document}